\nonstopmode

\newcommand{\wcmsq}{\ensuremath{\,{\rm W}\!/{\rm cm}^{2}} }
\newcommand{\csq}{$\langle\cos^{2}\theta\rangle$ }

\newcommand{\micron}{\ensuremath{\mu {\rm m}} }
\newcommand{\micronn}{\ensuremath{\mu {\rm m}}}
\newcommand{\cfbr}{${\rm CF}_{3}{\rm Br}$ }
\newcommand{\cfbrn}{${\rm CF}_{3}{\rm Br}$}

\documentclass[prb,showpacs,twocolumn]{revtex4}
\usepackage{graphicx}

\begin{document}
\title{An x-ray probe of laser-aligned molecules}
\author{Emily R.~Peterson}
\altaffiliation[Present address: ]{University of Michigan, Ann Arbor,
Michigan~48109, USA}
\author{Christian Buth}
\author{Dohn A.~Arms}
\author{Robert W.~Dunford}
\author{Elliot P.~Kanter}
\author{Bertold Kr\"assig}
\author{Eric C.~Landahl}
\author{Stephen T.~Pratt}
\author{Robin Santra}
\author{Stephen H.~Southworth}
\author{Linda Young}
\affiliation{Argonne National Laboratory, Argonne, Illinois~60439, USA}
\date{07 March 2008}

\begin{abstract}
We demonstrate a hard x-ray probe of laser-aligned small molecules.
To align small molecules with optical lasers, high intensities at nonresonant
wavelengths are necessary.
We use 95$\,$ps pulses focused to 40~\micron from an 800$\,$nm Ti:sapphire laser
at a peak intensity of~$10^{12}$ W/cm$^2$ to create an
ensemble of aligned bromotrifluoromethane~(\cfbrn) molecules.
Linearly polarized, 120$\,$ps x-ray pulses, focused to 10~\micronn, tuned to the
Br~$1s\rightarrow\sigma$*~pre-edge resonance at~13.476$\,$keV, probe the ensemble
of laser-aligned molecules.
The demonstrated methodology has a variety of applications and can enable
ultrafast imaging of laser-controlled molecular motions with
{\AA}ngstrom-level resolution.
\end{abstract}

\maketitle

Intense laser fields have greatly expanded our ability to control the behavior of
isolated atoms and molecules.~\cite{Yamanouchi,Seideman03}
A nonresonant, linearly polarized laser field will align a molecule by
interaction with the molecule's anisotropic polarizability;
the most polarizable axis within the molecule will align parallel to the
laser polarization axis.~\cite{Friedrich95}
Since the laser polarization direction is under simple
control with a waveplate, so is the direction of the molecule's most polarizable
axis with respect to the laboratory frame. The polarizability interaction used to
align molecules is identical to that used in optical
trap~\cite{Ash70PRL,Ash71APL} studies of biomolecules, though the alignment aspect is
not usually emphasized.~\cite{Neu04RSI}
Such experiments routinely use visible light probes and achieve nanometer-level
resolution.~\cite{Neu04RSI}
Here we demonstrate the use of 0.9~{\AA} x-rays to probe a transient ensemble of
laser-aligned molecules, thus, taking a step toward {\AA}ngstrom-level ultrafast
imaging of molecular motions.

Laser control of molecular alignment enables control over x-ray processes. For
instance, scattering from an ensemble of aligned molecules produces Bragg-like
diffraction spots rather than the concentric rings observed in scattering from an
isotropic gas.  An important application is biomolecule structure determination
with few {\AA}ngstrom resolution using coherent diffractive imaging with hard x-ray
free-electron lasers.~\cite{Neutze00}
The original concept~\cite{Neutze00} did not suggest aligned molecules but rather
proposed to scatter $10^{12}$ x-rays from a single biomolecule within 10$\,$fs
and collect a diffraction pattern with sufficient information to determine the
molecule's orientation in a single shot.
A recent experiment using the FLASH free-electron laser operating at 320$\,${\AA}
provides an important proof of principle.~\cite{Chapman07}
As an alternative, having prealigned molecules will vastly simplify the data
collection and analysis.~\cite{Spence04}
While both proposals~\cite{Neutze00,Spence04} focus on x-ray
scattering from a {\it single} large molecule and multiple repetition to build up
statistics, i.e., ``serial crystallography,"  our work focuses on x-ray probing of
an {\it ensemble} of~$10^7$ small molecules in the gas phase which have been
aligned with laser techniques.~\cite{Seideman03}
This strategy will allow one to acquire x-ray diffraction patterns of aligned,
noninteracting molecules and, thus, obtain {\AA}ngstrom-level molecular images
using existing synchrotron sources.

To demonstrate the essential ideas, we focus here on the simpler situation of
resonant x-ray absorption to probe laser-aligned molecules.
X-ray absorption resonances, resulting from the promotion of a $1s$ electron
(localized on a given atom) to an empty $\sigma$* or $\pi$* orbital (fixed to the
molecular frame), are sensitive to the relative angle between the molecular axis
and the x-ray polarization axis.~\cite{Stohr}
We studied the nonhazardous, symmetric top molecule bromotrifluoromethane~(\cfbrn).
Using {\it ab initio} methods, we calculate that the C--Br~axis is the most
polarizable and will align parallel to the laser polarization axis.
The \cfbr alignment axis can be rotated with respect
to the x-ray polarization axis and a linear dichroism indicative of the degree of
molecular alignment results.
For~\cfbrn, the Br$\,1s\rightarrow\sigma$* pre-edge
resonance is an excitation from the Br $1s$ orbital to an antibonding $\sigma$*
orbital containing substantial Br$\,4p_z$ character, where $z$ refers to the
C--Br~axis.
As a result of the molecular symmetry, x-ray absorption on the
Br$\,1s\rightarrow\sigma$* resonance occurs only when the x-ray polarization vector
has a nonvanishing projection on the C--Br~axis.

Two basic experiments demonstrate the x-ray probe of laser-aligned molecules.
First, we show that resonant x-ray absorption changes
reversibly in the presence of the laser field by measuring a laser/x-ray
cross correlation.
The ensemble is aligned only transiently and a theory
including time dependence is required to describe the dynamics of molecular
alignment.~\cite{Torres}
Second, we demonstrate control of x-ray absorption on the
Br$\,1s\rightarrow\sigma$*~resonance by rotating the alignment of~\cfbr
molecules with respect to the x-ray polarization axis.  For each of these
experiments, we compare with a theory developed to describe x-ray absorption of
laser-aligned molecules.~\cite{Buth}

\begin{figure}
  \centerline{\includegraphics [clip,width=3in, angle=0]{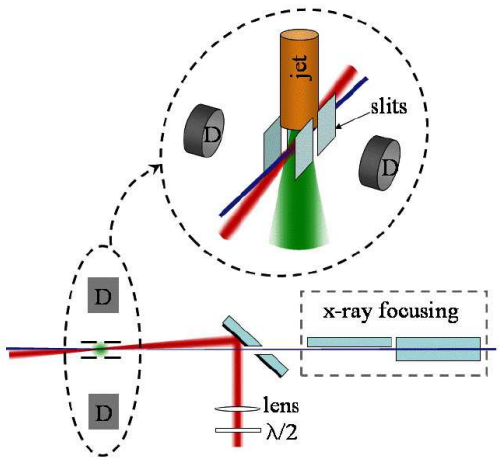}}
  \caption{(Color online) Schematic of experimental setup, top view.
           Inset with projection of interaction region: jet, collimation
           slit geometry and Si drift detectors~(D).}
  \label{FIGapparatus}
\end{figure}

Our experimental arrangement (Fig.~\ref{FIGapparatus}) used microfocused
x-rays to probe rotationally cooled \cfbr in the presence of the linearly
polarized aligning laser field. A $\lambda/2$ waveplate controlled
$\vartheta_{\mathrm{LX}}$, the angle between the x-ray and laser polarization
axes. Monochromatic, linearly polarized x-rays near 13.5$\,$keV [bandwidth
is~0.7$\,$eV full width at half maximum~(FWHM)] from an undulator source at
Sector~7 of the Advanced Photon Source were focused to 10$\,$\micron (FWHM) and
overlapped collinearly with the 40$\,$\micron (FWHM)
aligning laser beam.
Ti:sapphire oscillator pulses were stretched and amplified
to produce alignment pulses (1.9$\,$mJ, 95$\,$ps FWHM) at a repetition rate of 1$\,$kHz.
The four dimensional overlap (temporal and spatial) was done with methods
previously described.~\cite{Young}
The signature of \cfbr x-ray absorption, Br~$K\alpha$ fluorescence
at~11.9$\,$keV, was viewed with Si~drift detectors through molybdenum slits
that limited the detection to the central 1.2$\,$mm of the laser/x-ray overlap
region.
Spatial averaging over the crossing
angle and viewed region yielded a peak laser intensity of $(0.85\pm 0.09)
\times 10^{12}\,$\wcmsq.
\cfbr was rotationally cooled by expanding a mixture of
10\%~\cfbrn/90\% helium through pinhole nozzles of diameter $d = 25$ or $50\,\micron$
at backing pressures up to $P_{0}=9\,$bar.
The laser and x-ray beams intersected
the supersonic expansion $\sim 1\,$mm downstream of the nozzle, where the number
density of~\cfbr was  $\sim 5\times10^{14} \, $/cm$^3$.

\begin{figure}
  \centerline{\includegraphics [clip,width=3in, angle=0]{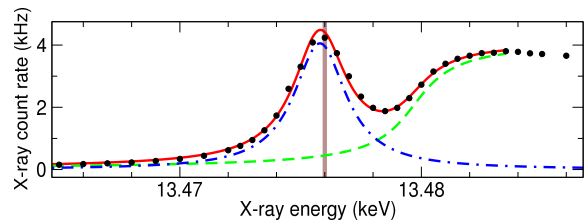}}
  \caption{(Color online) X-ray absorption spectrum of~\cfbr near the
           Br~{\em K}~edge.
           Experimental data (dots).
           The Br$\,1s\rightarrow\sigma$*
           resonance of~\cfbr at 13.476$\,$keV is marked with a vertical line.
           The fitted components and their sum are shown: Br$\,1s\rightarrow\sigma$*
           resonance (dash-dot), Br~{\em K}~edge (dashed), and sum (solid).}
  \label{K_edge}
\end{figure}

The x-ray absorption spectrum of~\cfbr shown in Fig.~\ref{K_edge} was obtained by
collecting Br~$K\alpha$ fluorescence as a function of the incident x-ray energy.
The Br$\,1s\rightarrow\sigma$* resonance at 13.476$\,$keV is the prominent feature
below the Br {\em K}-edge energy.  The resonance was fit to a Lorentzian with
$\Gamma=2.6\,$eV and the Br~{\em K}-edge plus Rydberg excitations were fit to an
arctangent function.~\cite{Agrawal}
The arctangent function contributes a 10\%
background under the Br$\,1s\rightarrow\sigma$*~resonance.

\begin{figure}
  \centerline{\includegraphics [clip,width=3in, angle=0]{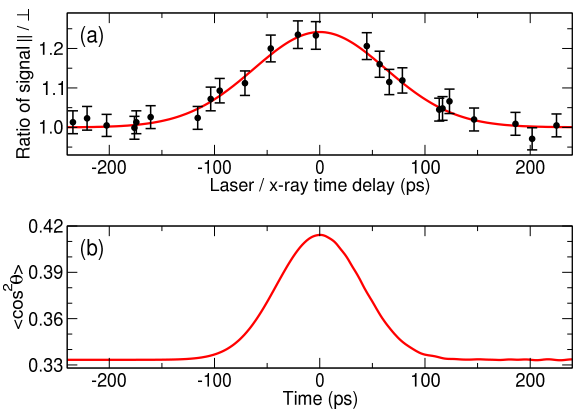}}
  \caption{(Color online) (a)~Laser/x-ray cross correlation for
           laser-aligned~\cfbrn, $\parallel/\perp$ absorption as a function
           of laser/x-ray delay.
           Theory for a rotational temperature~$T_{\mathrm{rot}}=20\,$K is
           overlaid.
           (b)~Calculated~\csq for a 95$\,$ps (FWHM) laser pulse with an intensity
           of~$0.85\times 10^{12} \, \wcmsq$ for $T_{\mathrm{rot}}=20\,$K.}
  \label{cos_cross}
\end{figure}

In the first experiment, we measured a laser/x-ray cross-correlation  signal by
tuning the x-ray energy to the Br $1s\rightarrow\sigma$*  resonance and varying
the laser/x-ray time delay~$\tau$.
Figure~\ref{cos_cross}(a) shows the cross-correlation signal, defined by the
ratio of parallel ($\vartheta_{\mathrm{LX}}=0^{\circ}$) to perpendicular
($\vartheta_{\mathrm{LX}}=90^{\circ}$)  x-ray absorption, as a function of
laser/x-ray delay.
The ratio was formed after subtracting the 10\% background
discussed above.
The alignment evolves adiabatically---following the laser pulse
envelope.
A Gaussian fit to the cross correlation signal yields a FWHM of $150\pm
14\,$ps and amplitude of $1.22 \pm 0.03$.
We calculated for comparison with experiment the absorption cross sections for
(1)~parallel laser and x-ray polarizations, (2)~perpendicular laser and x-ray
polarizations, and (3)~a laser-free thermal ensemble.~\cite{Buth}
In these calculations, the only adjustable parameters are the rotational temperature and
the pulse length of the x-rays.
The relevant molecular parameters, e.g., rotational constants, calculated by
theory are in agreement with measured values.
Comparison of the cross-correlation signal with theory yields an x-ray pulse
length of $122\pm18$~ps, consistent with expectations.

The usual criterion for adiabatic molecular response is
$\tau_{\mathrm{pulse}}\gg\tau_{\mathrm{rot}}$, where $\tau_{\mathrm{pulse}}$ is
the laser pulse duration and $\tau_{\mathrm{rot}}$ is the rotational period of the
molecule.~\cite{Seideman03}
A 20$\,$K thermal ensemble of~\cfbr has an average
rotational period of 28$\,$ps, much less than our 95$\,$ps laser pulse duration, thus,
fulfilling the adiabatic criterion (the ground state rotational period of
\cfbr is~235$\,$ps for rotation about an axis perpendicular to the C--Br
axis~\cite{Bvalue}).
The commonly used measure of molecular alignment is
$\langle\cos^{2}\theta\rangle$, where $\theta$ is the angle between the laser
polarization and the molecular axis.
For our experimental parameters, we calculated that the evolution of \csq follows
the laser pulse envelope, as shown in Fig.~\ref{cos_cross}(b).

\begin{figure}
  \centerline{\includegraphics[clip,width=3in]{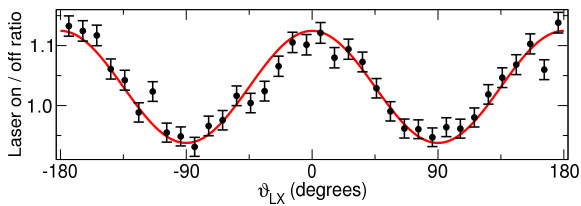}}
  \caption{(Color online) Laser on/laser off ratio of Br~fluorescence induced
           by x-ray excitation on the Br~$1s\rightarrow\sigma$* resonance
           of~\cfbr at 13.476$\,$keV, as a function of~$\vartheta_{\mathrm{LX}}$.
           $\vartheta_{\mathrm{LX}}=0^{\circ}$ and $90^\circ$ correspond to
           the parallel and perpendicular configurations for laser and x-ray
           polarizations.
           Theory for $T_{\mathrm{rot}}=24\,$K is the overlaid curve.}
  \label{waveplate}
\end{figure}

After optimizing temporal overlap at $\tau=0$,  we demonstrate control of resonant
x-ray absorption by varying the angle between the laser and x-ray polarizations,
$\vartheta_{\mathrm{LX}}$  (Fig.~\ref{waveplate}).
The quantity plotted on the ordinate is the ratio of
the Br $K\alpha$ fluorescence signal from the aligned sample (laser on) to the
isotropic sample (laser off).
A maximum occurs for
the parallel configuration $\vartheta_{\mathrm{LX}}=0^{\circ}$ and a
minimum for the perpendicular configuration $\vartheta_{\mathrm{LX}}=90^{\circ}$.
This confirms the symmetry of the $\sigma$* resonance with respect to the
C--Br~axis.
To compare with experiment, we calculated the $\vartheta_{\mathrm{LX}}$
dependence of the laser-on/laser-off ratio at $\tau=0$, as shown in Fig.~\ref{waveplate}
where good agreement is found for a rotational temperature
$T_{\mathrm{rot}}= 24 \pm 2\,$K and x-ray pulse duration of 122$\,$ps, determined from
the cross-correlation measurement.

Significant improvement in the degree of alignment can be achieved with lower
rotational temperature or higher laser alignment intensity.
At 20$\,$K, we observed a linear dependence of the alignment up to our maximum
peak intensity of $\sim 10^{12}\,$W/cm$^2$ with no ionization.
Our expansion provided significantly less rotational cooling than a pulsed
expansion ($P_{0}=100\,$bar, $d=200\,\micronn$)
where rotational temperatures $\sim 1.5\,$K have been obtained.~\cite{Kumar06}
Our calculations for 1$\,$K predict the alignment magnitude to increase to
\csq $=0.80$ and the alignment duration to 140$\,$ps
(FWHM) for $\tau_{\mathrm{pulse}}=95\,$ps.

The present work suggests several applications using x-ray absorption.
First, polarized x-ray absorption by molecules
whose axes are aligned in space enables the assignment of the symmetries (e.g.,
$\sigma$* and $\pi$*) of near edge resonances.  These techniques have been used
for molecules adsorbed on surfaces,~\cite{Stohr} but they can now be extended
generally to isolated molecules in the gas phase.  They are more general than
angle-resolved photoion yield spectroscopic methods~\cite{Adachi} which rely on
ion dissociation and the axial recoil approximation.
In addition, x-rays are ideal probes of laser-aligned molecules in the solution
phase, where Coulomb explosion imaging techniques typically used in gas phase
experiments~\cite{Seideman03,Kumar06} are not applicable.
An x-ray probe is also free from the complications of strong-field probes arising
from complex ionization pathways.~\cite{Pavicic07}
There also is an opportunity to use x-rays as a probe
of laser-controlled rotations to achieve a better understanding of solvent
dynamics.~\cite{Ramakrishna05}
The penetrating power of x-rays in low-$Z$ solvents,
combined with the localized core-level absorption on the solute molecule, are both
significant assets.  The advent of tunable, polarized 1$\,$ps x-rays at the Advanced
Photon Source~\cite{Borland05} will enable tracking rotational dynamics in
condensed phases as well as impulsively aligned molecules in the gas phase.

Finally, our work demonstrates column densities of aligned molecules that are
sufficient for coherent diffraction imaging.
We already have achieved an aligned molecule density of~$\sim 5\times10^{14}
\,$/cm$^3$, corresponding to $4\times10^7$ molecules within the laser/x-ray
overlap volume.
With this density, we estimate that less than $6\,$min is required to acquire
a coherent diffraction image for the prototypical Br$_2$ with a total of $10^5$
events, assuming an elastic scattering cross section of 1.6$\,$kb (obtained by
summing the contributions of the individual atoms), $10^8$
x-rays/pulse/$(10\,\micronn)^2$ at a repetition rate of 1$\,$kHz.
At a $10\%$ seeding fraction, the coexpanded He would provide only a $\sim
2\%$ background and dimer formation is expected to be $<1\%$.
Thus, this demonstration can be considered a first step toward {\AA}ngstrom-level
x-ray imaging of uncrystallized molecules.

\begin{acknowledgments}
This work and the Advanced Photon Source were supported by the Chemical Sciences,
Geosciences, and Biosciences Division  of the Office of Basic Energy Sciences,
Office of Science, U.S.~Department of Energy, under Contract
No.~DE-AC02-06CH11357.
C.B.~was partly supported by a Feodor Lynen Research Fellowship
from the Alexander von Humboldt Foundation.
\end{acknowledgments}

\end{document}